\documentclass[12pt]{article}

\usepackage{graphicx, amsmath,amsfonts,amssymb}
\usepackage{authblk}

\newcommand{\ra}{\rightarrow}
\newcommand{\minus}{\backslash}
\newcommand{\ZZ}{{\mathbb Z}}
\newcommand{\RR}{{\mathbb R}}
\newcommand{\CC}{{\mathbb C}}

\newcommand{\cV}{{\mathcal V}}

\newcommand{\cA}{{\mathcal A}}
\newcommand{\cE}{{\mathcal E}}

\newcommand{\dist}{{\rm dist}}

\begin{document}

\title{Local Commuting Projector Hamiltonians and the Quantum Hall Effect}

\author{Anton Kapustin}
\affil{{\it California Institute of Technology, Pasadena, California }}
\author{Lukasz Fidkowski}
\affil{{\it Department of Physics, University of Washington, Seattle}}

\maketitle

\abstract{We prove that neither Integer nor Fractional Quantum Hall Effects with nonzero Hall conductivity are possible in gapped systems described by Local Commuting Projector Hamiltonians.}

\newpage

\section{Introduction}

One of the simplest classes of exactly soluble many-body Hamiltonians is the class of Local Commuting Projector Hamiltonians (LCPH). These are finite-range lattice Hamiltonians which have the form
$$
H=\sum_\Gamma \Phi(\Gamma),
$$
where the local terms $\Phi(\Gamma)$ are commuting projectors. Usually one also assumes that each local Hilbert space is finite-dimensional, otherwise one gets a hugely degenerate excitation spectrum.  The toric code \cite{toriccode}, or more generally, Levin-Wen Hamiltonians associated to unitary spherical fusion categories \cite{LevinWen}, provide interesting examples of such Hamiltonians in two dimensions, so many topologically ordered states can be described by LCPH. It is believed that all known Symmetry Protected Topological phases of fermions and bosons with a finite symmetry can also be described by LCPH.  Nevertheless, it is also widely believed that neither IQHE nor FQHE phases can be realized by LCPH. In this short note, we provide a proof of this. 

More precisely, we prove the following. Suppose $H$ is a local lattice Hamiltonian with an on-site $U(1)$ symmetry defined on a torus of size $L$. It is well-known that it is possible to extend $H$ to a 2-parameter family $H(\beta_x,\beta_y)$ of local lattice Hamiltonians depending on the ``holonomies''\footnote{In many papers ``holonomies'' are called fluxes. We find this terminology confusing, since the word ``flux" is also used to describe a region of nonzero magnetic field, while $\beta_x,\beta_y$ parameterize a flat $U(1)$ gauge field on a torus.} $(\beta_x,\beta_y)\in \RR^2/(2\pi\ZZ)^2$. If $H(\beta_x,\beta_y)$ has a unique ground state for all $\beta_x,\beta_y$, then its ground-states form a rank-one vector bundle $E$ over $T_\beta=\RR^2/(2\pi\ZZ)^2$. It was argued in  \cite{ThoulessNiu,AvronSeiler} that the first Chern number of $E$ is equal to the Hall conductance of the system. More precisely, the Hall conductance is defined in the thermodynamic limit $L\ra\infty$. If one assumes that the limiting ground state exists, and the spectral gap does not close in the limit $L\ra\infty$, one can indeed prove that the thermodynamic limit of the first Chern number of $E$ exists and is equal to the Hall conductance \cite{hall1,hall2,hall3}. (There is an alternative proof of the quantization of the Hall conductance which only requires $H$ to be gapped, but does not make any assumption about the gap for nonzero $\beta_x,\beta_y$ \cite{HM}). This line of reasoning extends to the case when the ground-state is degenerate \cite{hall2}: if the thermodynamic limit of all ground-states is the same, then the limit of the first Chern number exists and is equal to $q$ times the Hall conductance, where $q$ is the degeneracy of the ground-states on a torus of a sufficiently large size. We prove the following

{\bf Theorem.}
{\it Let $H$ be an LCPH with range $R$ on a torus of size $L>4R$, and suppose $H$ has an on-site $U(1)$ symmetry. Then the 2-parameter family $H(\beta_x,\beta_y)$ of Hamiltonians depending on the "holonomies" $\beta_x,\beta_y\in \RR^2/(2\pi\ZZ)^2$ is a family of LCPH, and thus the gap assumption is satisfied. Moreover, the Chern number of the corresponding bundle of ground-states vanishes.}

Assuming that the thermodynamic limit exists, this implies that neither IQHE nor FQHE states can be realized by Local Commuting Projector Hamiltonians with an on-site $U(1)$ symmetry.

While for simplicity we only discuss the case $d=2$, in arbitrary dimension the same arguments show that all Chern classes of the bundle of ground-states vanish if $H$ is an LCPH, and in fact the bundle of ground-states is topologically trivial. 

The proof uses some well-known results from algebraic geometry. The same mathematical results were used in \cite{BK} to show that a Chern insulator with a finite-range band Hamiltonian  cannot have correlations which decay faster than any exponential. This is no coincidence: correlations of all local observables in a ground-state of an LCPH vanish beyond a finite range. One might conjecture that neither IQHE nor FQHE can occur if correlations of all local observables decay faster than any exponential.

The content of the paper is as follows. In section 2, we recall the definition of a local lattice Hamiltonian on a torus with holonomies following \cite{HM,hall2}. The proof of the theorem occupies sections 3 and 4. 

The work was partly performed at the Aspen Center for Physics, which is supported by National Science Foundation grant PHY-1607611.
The research of A.\ K.\ was supported by the U.S.\ Department of Energy, Office of Science, Office of High Energy Physics, under Award Number DE-SC0011632 and by the Simons Investigator Award.  L.\ F.\ was supported by NSF DMR-1519579

\section{Lattice Hamiltonian on a torus with holonomies}

The space will be a torus $T^2$ of size $L\times L$. We identify it with $\RR^2/(L\ZZ)^2$. A lattice  is a finite subset $\Lambda\subset T^2$. The Hilbert space is either a tensor product 
$$
\cV=\otimes_{\lambda\in\Lambda} \cV_\lambda,
$$
where $\cV_\lambda$ is a finite-dimensional Hilbert space, or a super-tensor product
$$
\cV=\widehat\otimes_{\lambda\in\Lambda} \cV_\lambda,
$$
where $\cV_\lambda$ is a finite-dimensional $\ZZ_2$-graded Hilbert space. We will denote by $\cA_\Gamma$ the algebra of observables supported at a subset $\Gamma\in\Lambda$. Following \cite{HM}, we use an $\ell^1$ distance on $T^2$:
$$
\dist(p,p')=|x(p)-x(p')| {{\rm mod} L}+|y(p)-y(p')| {{\rm mod} L}.
$$

A local lattice Hamiltonian with range $R$  has the form
$$
H=\sum_{\Gamma} \Phi(\Gamma),
$$
where $\Phi(\Gamma)$ is a Hermitian element of $\cA_\Gamma$, and the sum is over all  subsets $\Gamma\subset\Lambda$ of diameter less than $R$. In the $\ZZ_2$-graded case, each $\Phi(\Gamma)$ is required to be even. In addition, one usually assumes that the norms of the operators $\Phi(\Gamma)$ are uniformly bounded \cite{HM}. In the cases of interest to us, this will be automatic. 

A local lattice Hamiltonian is said to have an on-site symmetry $G$ if each $\cV_\lambda$ is a unitary (or anti-unitary) representation of $G$, and each $\Phi(\Gamma)$ commutes with the action of $G$ on $\cA_\Gamma$. In particular, a local lattice Hamiltonian has an on-site symmetry $U(1)$ if for each $\lambda\in\Lambda$ we are given an (even) Hermitian operator $Q_\lambda:\cV_\lambda\ra \cV_\lambda$ with integral eigenvalues,  and for all $\Gamma$ of diameter less than $R$ we have
$$
[\Phi(\Gamma),Q(\Gamma)]=0,
$$
where 
$$
Q(\Gamma)=\sum_{\lambda\in\Gamma} Q_\lambda.
$$
Note that $Q(\Gamma)$ is additive under disjoint union:
$$
Q(\Gamma\cup\Gamma')=Q(\Gamma)+Q(\Gamma'),\quad {\rm if}\ \Gamma\cap\Gamma'=\emptyset.
$$

\begin{figure*}
\centering
\includegraphics[scale=.4,trim={0.0cm 2cm 0cm 2cm},clip]{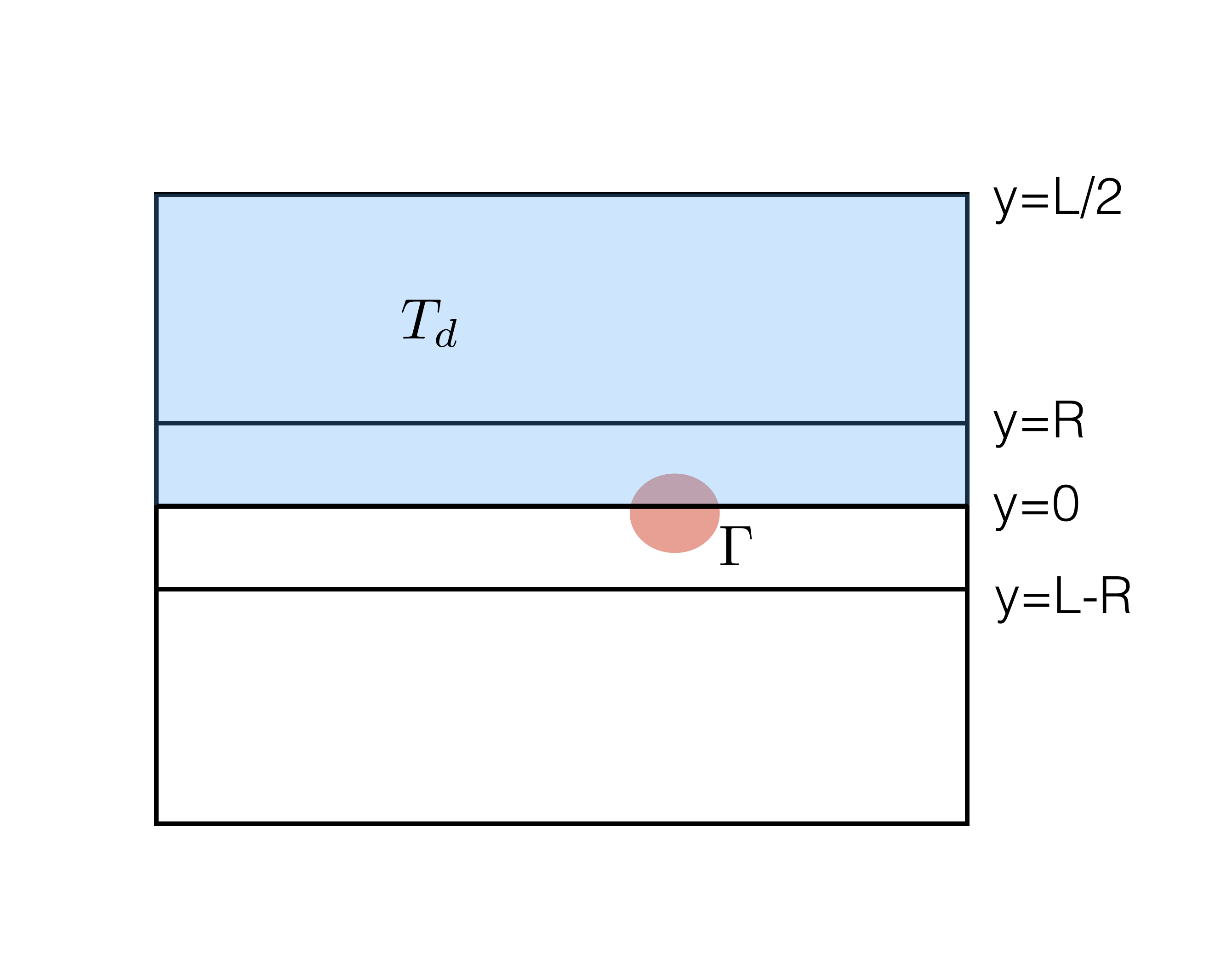}
\caption{The operator $\Phi(\Gamma,0,\beta_y)$ is equal to $\Phi(\Gamma)$ unless $\Phi(\Gamma)$ straddles the line $y=0$ (red region), in which case it is defined by $e^{-i\beta_y Q(\Lambda_d)} \Phi(\Gamma) e^{i\beta_y Q(\Lambda_d)}$, where $Q(\Lambda_d)$ is the operator that measures the $U(1)$ charge of $T_d$ (blue region).}
\label{fig:holonomy}
\end{figure*}

Given a local lattice Hamiltonian with an on-site $U(1)$ symmetry and range $R$, and assuming $L>2R$, one can define a family of local lattice Hamiltonians with range $R$ depending on $(\beta_x,\beta_y)\in \RR^2/(2\pi\ZZ)^2$ as follows \cite{HM,hall2,hall3}. Let $T_\ell$ be the subset of $T^2$ given by $0< x(p)< L/2$. Let $T_d$ be the subset of $T^2$ given by $0<y<L/2$. Let $\Lambda_\ell=\Lambda\bigcap T_\ell$ and $\Lambda_d=\Lambda\bigcap T_d$. We define (see Fig. \ref{fig:holonomy})
$$
\Phi(\Gamma,0,\beta_y)=\left\{ \begin{array}{ll} e^{-i\beta_y Q(\Lambda_d)} \Phi(\Gamma) e^{i\beta_y Q(\Lambda_d)} & {\rm if}\ \Gamma\bigcap\Lambda_d\neq \emptyset, \dist(\Gamma,\{y=0\})<R, \\ \Phi(\Gamma) & {\rm otherwise} \end{array}\right.
$$
Then we define 
$$
\Phi(\Gamma,\beta_x,\beta_y)=\left\{ \begin{array}{ll} e^{-i\beta_x Q(\Lambda_\ell)} \Phi(\Gamma,0,\beta_y) e^{i\beta_x Q(\Lambda_\ell)} & {\rm if}\ \Gamma\bigcap\Lambda_\ell\neq \emptyset, \dist(\Gamma,\{x=0\})<R, \\ \Phi(\Gamma,0,\beta_y) & {\rm otherwise} \end{array}\right.
$$
It follows from $U(1)$-invariance of $\Phi(\Gamma)$ and additivity of $Q(\Gamma)$ that $\Phi(\Gamma,\beta_x,\beta_y)$ is an element of $\cA_\Gamma$, and thus
$$
H(\beta_x,\beta_y)=\sum_\Gamma \Phi(\Gamma,\beta_x,\beta_y)
$$
is a local lattice Hamiltonian. Since $Q_\lambda$ has integral eigenvalues, it is clear that each $\Phi(\Gamma,\beta_x,\beta_y)$ is $2\pi$-periodic in both $\beta_x$ and $\beta_y$. Thus $H(\beta_x,\beta_y)$ is a family of local lattice Hamiltonians parameterized by points of a torus $T_\beta=\RR^2/(2\pi\ZZ)^2$. 

Suppose the gap between the lowest and next-to-lowest eigenvalues  of $H(\beta_x,\beta_y)$ is strictly greater than zero for all $\beta_x,\beta_y$. Then the image of the projector to the eigenspace with the lowest eigenvalue is a well-defined smooth vector bundle over $T_\beta$ which we denote $E$. 

\section{Local Commuting Projector Hamiltonians}

A local commuting projector Hamiltonian (LCPH) is a local lattice Hamiltonian such that each $\Phi(\Gamma)$ is a projector ($\Phi(\Gamma)^2 = \Phi(\Gamma)$) and $[\Phi(\Gamma),\Phi(\Gamma')]=0$ for all $\Gamma,\Gamma'$. In this section we show that if an LCPH $H$ with range $R$ has an on-site $U(1)$ symmetry, then for $L>4R$ the Hamiltonian $H(\beta_x,\beta_y)$ is also an LCPH.

 It is obvious that all operators $\Phi(\Gamma,\beta_x,\beta_y)$ are projectors, so we just need to check that they all commute. The gist of this argument is that for any two fixed $\Gamma, \Gamma'$, one can find a common local $U(1)$ rotation that conjugates $\Phi(\Gamma,\beta_x,\beta_y)$ and $\Phi(\Gamma,\beta_x,\beta_y)$ into $\Phi(\Gamma)$ and $\Phi(\Gamma')$, respectively.  More formally, to check that $\Phi(\Gamma,0,\beta_y)$ commutes with $\Phi(\Gamma',0,\beta_y)$, it is sufficient to check this when $\Gamma\bigcap\Lambda_d\neq\emptyset$  and $\dist(\Gamma,\{y=0\})<R$. If $\Gamma'\bigcap\Lambda_d=\emptyset$, then $\Phi(\Gamma',0,\beta_y)=\Phi(\Gamma')$ commutes with both $\Phi(\Gamma)$ and $Q(\Lambda_d)$, and therefore commutes with $\Phi(\Gamma,0,\beta_y)$. Suppose now that $\Gamma'\bigcap\Lambda_d\neq\emptyset$. We have two cases: $\dist(\Gamma',\{y=0\})<R$ or $\dist(\Gamma',\{y=0\})\geq R$. In the former case, both $\Phi(\Gamma,0,\beta_y)$ and $\Phi(\Gamma',0,\beta_y)$ are conjugate to $\Phi(\Gamma)$ and $\Phi(\Gamma')$ by means of the same unitary operator $\exp(i\beta_y Q(\Lambda_d))$. Since $\Phi(\Gamma)$ and $\Phi(\Gamma')$ commute,  $\Phi(\Gamma,0,\beta_y)$ and $\Phi(\Gamma',0,\beta_y)$ also commute. In the latter case, since the diameter of $\Gamma'$ is less than $R$, and $R<L/4$, all points of $\Gamma'$ satisfy $R\leq y<3L/4$. So if we define $\tilde\Lambda_d\subset\Lambda$ by the condition $0<y<3L/4$, then $\Phi(\Gamma)$ commutes with $Q(\tilde\Lambda_d\minus\Lambda_d)$, and thus 
$$
\Phi(\Gamma,0,\beta_y)=e^{-i\beta_y Q(\tilde\Lambda_d)} \Phi(\Gamma) e^{i\beta_y Q(\tilde\Lambda_d)}.
$$
Since $\Gamma'\subset \tilde\Lambda_d$, $U(1)$-invariance obviously implies
$$
\Phi(\Gamma',0,\beta_y)=\Phi(\Gamma')=e^{-i\beta_y Q(\tilde\Lambda_d)} \Phi(\Gamma') e^{i\beta_y Q(\tilde\Lambda_d)}.
$$
Since both $\Phi(\Gamma)$ and $\Phi(\Gamma')$ commute, the same applies to $\Phi(\Gamma,0,\beta_y)$ and $\Phi(\Gamma',0,\beta_y)$.

Now that we know that $\Phi(\Gamma,0,\beta_y)$ and $\Phi(\Gamma',0,\beta_y)$  commute for all $\Gamma,\Gamma'$, we can use  the same argument with $\beta_x$ and $\Lambda_\ell$ instead of $\beta_y$ and $\Lambda_d$ to show that $\Phi(\Gamma,\beta_x,\beta_y)$ and $\Phi(\Gamma',\beta_x,\beta_y)$ commute for all $\Gamma,\Gamma'$. Thus $H(\beta_x,\beta_y)$ is an LCPH. 
The eigenvalues of such a Hamiltonian are non-negative integers. Given an integer $N$, eigenvectors are non-zero solutions of the equations
$$
\Phi(\Gamma,\beta_x,\beta_y)\vert\Psi\rangle=n_\Gamma\vert\Psi\rangle,
$$
where the numbers $n_\Gamma$ take values in $\{0,1\}$ and satisfy 
$$
\sum_\Gamma n_\Gamma=N.
$$
Ground-states are eigenvectors corresponding to the smallest possible $N$.\footnote{It is usually assumed that the ground-states have $N=0$, and thus all $n_\Gamma$ vanish. We allow for more general possibilities.} If this value is $N_0$, then the next smallest eigenvalue is at least $N_0+1$, and thus the gap condition is satisfied. This proves the first part of the theorem.

\section{The vanishing of the Chern number}

We are ready to show that any local commuting projector Hamiltonian which has  an on-site $U(1)$ symmetry has a vanishing ground-state Chern number for $L>4R$. Let the smallest eigenvalue of $H(\beta_x,\beta_y)$ be $N_0$, and the corresponding bundle of ground-states be $E$. We have seen above that this bundle decomposes as a direct sum
$$
E=\oplus_{\{n_\Gamma\}} E_{\{ n_\Gamma\}},\quad \sum_\Gamma n_\Gamma=N_0,
$$
where $n_\Gamma\in\{0,1\}$ is an eigenvalue of $\Phi(\Gamma,\beta_x,\beta_y)$. We will show that each $E_{\{n_\Gamma\}}$ is topologically trivial, which will imply that $E$ is trivial too.

Let $P_n(\Gamma,\beta_x,\beta_y)$, $n\in\{0,1\}$, be the projector to the eigenspace of $\Phi(\Gamma,\beta_x,\beta_y)$ with eigenvalue $n$:
$$
P_n(\Gamma,\beta_x,\beta_y)=\left\{ \begin{array}{ll} 1-\Phi(\Gamma,\beta_x,\beta_y), & n=0\\ \Phi(\Gamma,\beta_x,\beta_y), & n=1 \end{array} \right.
$$
Then the projector to $E_{\{n_\Gamma\}}$ is given by
$$
P_{\{n_\Gamma\}}(\beta_x,\beta_y)=\prod_\Gamma P_{n_\Gamma}(\Gamma,\beta_x,\beta_y). 
$$ 
Since each $Q(\Gamma)$ has integral eigenvalues, the matrix elements of $\Phi(\Gamma,\beta_x,\beta_y)$ are trigonometric polynomials in $\beta_x,\beta_y$.  Therefore the same is true about the  projector $P_{\{n_\Gamma\}}(\beta_x,\beta_y)$.

Trigonometric polynomials on $T_\beta$ can be holomorphically extended to its complexification $\CC^*\times\CC^*$, where $\CC^*=\CC\minus\{0\}$. Such a holomorphic extension is a Laurent polynomial on $\CC^*\times\CC^*$, i.e. an expression of the form
$$
\sum_{|m_x|<N,|m_y|<N} a_{m_x,m_y} z_x^{m_x} z_y^{m_y},
$$
where $z_x=e^{i\beta_x}$, $z_y=e^{i\beta_y}$. When we extend $P_{\{n_\Gamma\}}$ to a Laurent polynomial on $\CC^*\times\CC^*$, the coefficients are matrices (operators on the finite-dimensional Hilbert space $\cV$). 

Laurent polynomials can be thought of as regular algebraic functions on an algebraic torus $\CC^*\times\CC^*$. A projector whose entries are Laurent polynomials thus defines an algebraic vector bundle on $\CC^*\times\CC^*$ (by taking its image). Let us call $\cE_{\{n_\Gamma\}}$ the algebraic vector bundle corresponding to the projector $P_{\{n_\Gamma\}}$.  The bundle  $E_{\{n_\Gamma\}}$ is a restriction of $\cE_{\{n_\Gamma\}}$ to the real slice $T_\beta\subset\CC^*\times\CC^*$.  Now we can use the result from algebraic geometry \cite{Gubeladze,Lam} that any algebraic vector bundle over $(\CC^*)^d$ is isomorphic to a trivial vector bundle, for any $d$. Therefore $E_{\{n_\Gamma\}}$ and $E$ are trivial, which proves the second part of the theorem. One can also use a more elementary argument briefly explained in \cite{BK}: the first Chern class of $\cE_{\{n_\Gamma\}}$ is the same as the first Chern class of the line bundle $\Lambda^r \cE_{\{n_\Gamma\}}$, where $r$ is the rank of $\cE_{\{n_\Gamma\}}$ and the latter bundle is necessarily trivial because the ring of regular  algebraic functions on $\CC^*\times\CC^*$ is a Unique Factorization Domain. Or one can use an even more elementary deformation argument which directly shows that the Chern classes of any algebraic vector bundle on $\CC^*\times\CC^*$ are trivial \cite{DF}. 

Note that the numbers $n_\Gamma$ label the eigenvalues of local integrals of motion $\Phi(\Gamma)$. In the presence of such integrals of motion, the thermodynamic limit of a state, if it exists, may depend not only on the energy, but also on $n_\Gamma$. If this happens, then the Chern number of $E$ is not directly related to the Hall conductance. However, it is natural to assume that the thermodynamic limit of a state is completely characterized by the eigenvalues of local integrals of motion $n_\Gamma$. Then the Hall conductance is proportional to the Chern number of $E_{\{n_\Gamma\}}$. Since we showed that it vanishes for $L>4R$, we can still conclude that the Hall conductance vanishes.

\end{document}